\documentclass[aps,prb,twocolumn,superscriptaddress,showpacs]{revtex4}
\usepackage[latin1]{inputenc}
\usepackage{graphicx}
\usepackage{amsmath,amsfonts}

\begin{document}

\title{Enhanced transmission and beaming of light via photonic crystal
surface modes}

\author{Esteban Moreno}
\email[Electronic address: ]{esteban.moreno@uam.es}
\affiliation{Laboratory for Electromagnetic Fields and Microwave
Electronics, Swiss Federal Institute of Technology, ETH-Zentrum,
CH-8092 Zurich, Switzerland}

\affiliation{Departamento de Física Teórica de la Materia
Condensada, Universidad Autónoma de Madrid, E-28049 Madrid, Spain}

\author{F. J. García-Vidal}
\affiliation{Departamento de Física Teórica de la Materia
Condensada, Universidad Autónoma de Madrid, E-28049 Madrid, Spain}

\author{L. Martín-Moreno}
\affiliation{Departamento de Física de la Materia Condensada,
Universidad de Zaragoza-CSIC, E-50009 Zaragoza, Spain}

\begin{abstract}
Surface modes are generally believed to be an undesirable feature
of finite photonic crystals (PC), unlike point or line defect
modes. However, it is possible to make the surface mode radiate by
appropriate corrugation of the PC interface. In this paper we show
theoretically that the coherent action of these surface
indentations can be engineered to collimate within a few degrees
the light exiting a PC waveguide, or to funnel light coming from
free space into the waveguide.
\end{abstract}


\maketitle

The proposal of photonic crystals
(PC)~\cite{Yablonovitch87,John87} in dielectric materials has
stimulated the search of photonic band gap effects in periodic
metallic
structures~\cite{McGurn93,Smith94,Kitson96,FJ96,Fan96,Sievenpiper98,Bozhe01,Fleming02}.
These latter structures have shown remarkable properties, such as
extraordinary optical transmission both through subwavelength hole
arrays~\cite{Ebbesen98} and single apertures~\cite{Thio01}, and
beaming of light from single apertures flanked by periodic surface
corrugations~\cite{Lezec03}. The root of all these phenomena lies
in the existence of surface electromagnetic (EM) modes at
metal-dielectric interfaces~\cite{LMM01,LMM03,FJ03}. PCs can also
support surface EM modes for appropriate crystal termination
surfaces~\cite{Meade91,winn94,Ramos99}. Additionally, for
frequencies within the band gap, the PC behaves as a mirror for
incoming EM radiation. These two facts suggest that the cited
optical phenomena found in metals could have counterparts in PC
systems. However, this analogy is not straightforward as the field
penetration inside PCs (of the order of one wavelength) is much
larger than the skin depth in good metals in the optical regime,
which could strongly modify the interference phenomena we are
referring to. In this paper we explore the possibility of
exporting some of the results found in metallic structures to
dielectric PCs. We will show that, through appropriate corrugation
of the PC surface, it is possible to obtain both enhanced
transmission through slits in PC slabs, and strong beaming of
light coming out of a PC waveguide.

Let us first discuss the basics of the beaming phenomena in
metallic systems, considering the simpler one-dimensional case of
a guided mode coming out of one single slit flanked by a finite
array of grooves~\cite{LMM03}. The grooves act as centers that
re-scatter the EM radiation diffracted by the slit exit. For
certain resonant conditions a leaky surface EM wave develops,
which emits into vacuum as it propagates along the surface. The
phase accumulated during this surface propagation can be
compensated, after emission, by optical paths in free space.
Therefore, for certain angles, radiation coming from all
indentations interferes constructively, leading to beaming. This
mechanism suggests the strategy shown in this paper in order to
pursue the analogy: first a waveguide in a PC is constructed and
then, by modifying the PC-vacuum interface, a surface EM mode is
created. Finally, in order to achieve the desired beaming effect,
this mode is connected to the radiative modes by adding a periodic
modulation to the PC surface. Additionally, these surface leaky
modes could also be used to funnel light from free space into the
waveguide, therefore enhancing the transmission, as has been
previously demonstrated in single apertures surrounded by a
periodic corrugation in a metallic film~\cite{FJ03}.

\begin{figure*}
\includegraphics{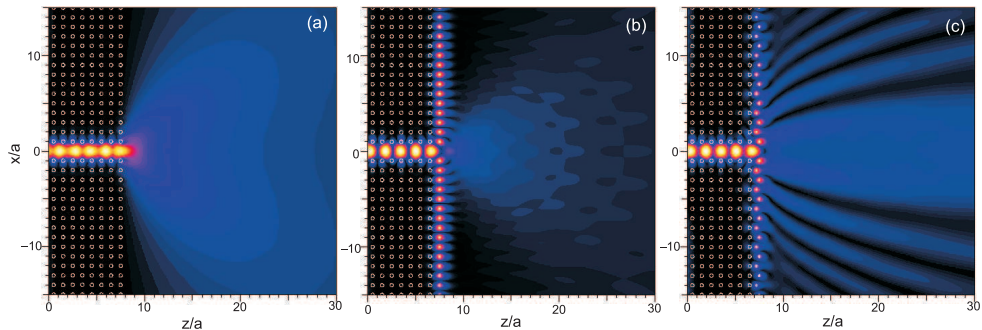}
\caption{\label{electric_field}Electric field amplitude $|E(z,
x)|$ radiated out of a PC waveguide for various crystal
interfaces. The input field is a guided mode coming from the left.
(a)~Standard interface (no surface modes allowed), (b)~the
cylinders at the interface have smaller radius than those in the
PC bulk (one non-radiative surface mode supported), and (c)~same
as in (b) but with additional periodic modulation of the interface
cylinders (one leaky surface mode supported). The color scales are
linear and extend from black (0) to white (maximum), the units are
arbitrary.}
\end{figure*}

\begin{figure}[b]
\includegraphics{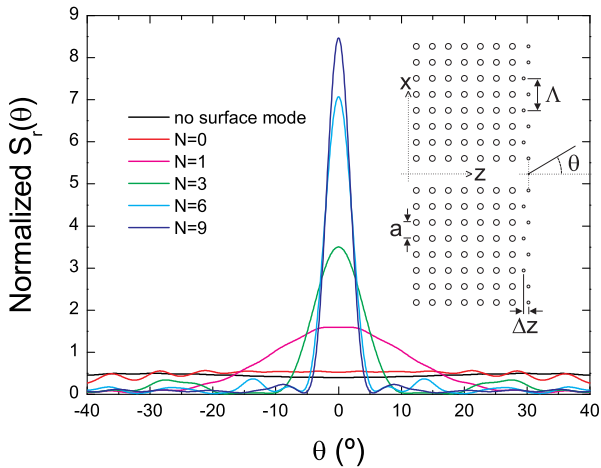}
\caption{\label{radiation_pattern}Far-field radial component of
the Poynting vector, $S_r$, radiated out of a PC waveguide as a
function of azimuthal angle, $\theta$. In all cases, the integral
in $\theta$ of the angular transmission distribution, $r
S_r(\theta)$, is normalized to unit. The black curve corresponds
to a standard interface, whereas the remaining curves correspond
to an interface with cylinders radius $r_\textrm{i}= 0.09 a$. $N$
stands for the number of corrugations in the modulated surface
(see text). Inset: scheme of the considered structures.}
\end{figure}

\begin{figure*}
\includegraphics{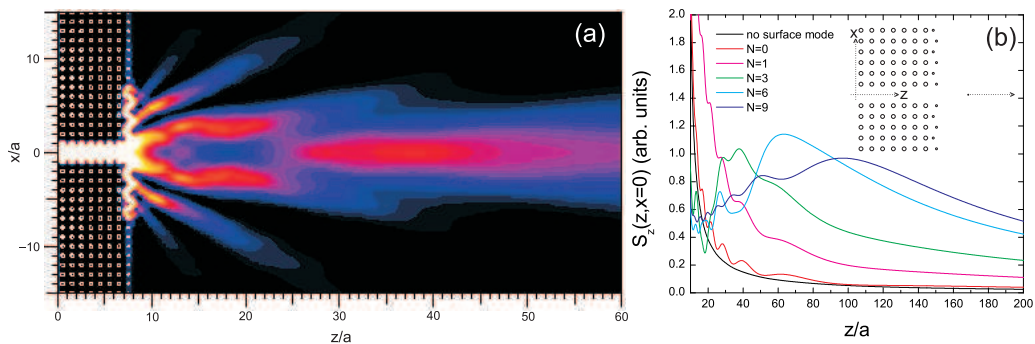}
\caption{\label{axial_cuts}(a)~Poynting field modulus $S(z, x)$
radiated out of a PC waveguide. The PC-vacuum interface supports
one surface mode which radiates due to a periodic modulation of
the surface (number of corrugations $N=3$) leading to beaming and
lensing. (b)~Poynting field $S_z(z, x=0)$ along the waveguide axis
(the dotted line in the inset) for increasing number of
corrugations at the interface.}
\end{figure*}

\begin{figure}[b]
\includegraphics{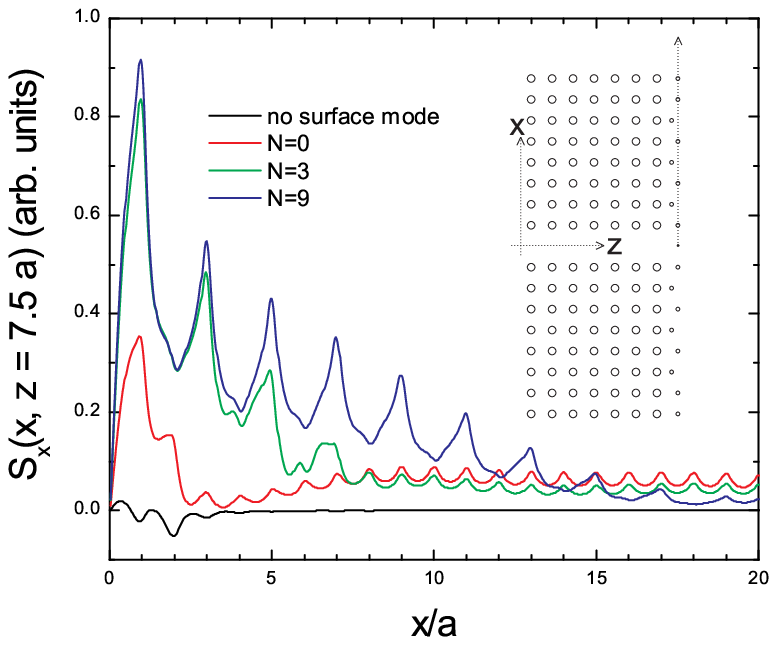}
\caption{\label{interface_cuts}Poynting field $S_x(x, z=7.5a)$
along the PC-vacuum interface (dotted line in the inset). The
black curve corresponds to a standard interface (no surface mode
supported), whereas the remaining curves correspond to an
interface with cylinders radius $r_\textrm{i}= 0.09 a$ (one
surface mode supported). $N$ stands for the number of corrugations
in the modulated surface.}
\end{figure}

In this paper, for proof of principle purposes, we concentrate in
two-dimensional PCs. As photonic band gap (PBG) system, we
consider a square lattice of cylinders in vacuum. We choose
cylinder radius $r=0.18 a$ ($a$ being the lattice parameter) and
dielectric constant of the cylinders $\epsilon_\textrm{r}=11.56$
(we take $\epsilon_\textrm{r}$ for GaAs at a wavelength of
$1.5\,\mu m$ as a representative value). For this set of
parameters the system presents, for E-polarization (electric field
pointing along the cylinders axis), a frequency gap for reduced
frequencies $\omega a/2 \pi c$ in the range $(0.30, 0.44)$, where
$c$ is the speed of light. All results presented are for
E-polarization and have been obtained with the multiple multipole
(MMP) method~\cite{Hafner99}. The computational technique to treat
PC waveguide discontinuities within the MMP method is thoroughly
described in Refs.~\onlinecite{Moreno02a,Moreno02b}. A waveguide
is constructed by removing one row of cylinders in the
$z$-direction (see inset of Fig.~\ref{radiation_pattern}). For
most frequencies in the PBG this waveguide supports one
propagating mode. The PC is infinitely extended towards
$z\rightarrow-\infty$ but, in order to study the coupling of the
waveguide mode into free space, we terminate the PC in a set of
cylinders parallel to the $x$-direction (in the coordinate system
we have chosen, the center of these last cylinders is at $z=7.5
a$). This regular PC termination does not support surface
states~\cite{winn94}. Fig.~\ref{electric_field}(a) shows the real
space electric field amplitude when a guided mode with frequency
$\omega_0= 0.408 (2 \pi c/a)$, launched at $z\rightarrow-\infty$,
is diffracted by the waveguide exit. The radiated electric field
shows a broad angular distribution, as corresponds to an aperture
of dimensions comparable to the wavelength (the wavelength
corresponding to $\omega_0$ is $\lambda_0=2.45 a$). This is more
clearly seen in Fig.~\ref{radiation_pattern} (black line),
rendering the far-field radial component of the Poynting vector,
$S_r$, as a function of azimuthal angle, $\theta$. In order to
create a surface state, one possibility would be to cut the outer
part of the interface cylinders, thus leaving one monolayer of
hemicylinders~\cite{winn94}. The origin of this surface state is
that a mode that would be in the dielectric band for complete
cylinders shifts its energy up when part of the dielectric is
removed~\cite{Meade91}. In this work the surface state is induced
in a similar manner by decreasing the radius $r_\textrm{i}$ of the
interface monolayer of cylinders, which is a more convenient
system to treat by the MMP method. For the ulterior illustration
of beaming in the forward direction, it is convenient to choose
$r_\textrm{i}$ in such a way that a surface mode appears with a
modal wavelength $\lambda_{\textrm{sm}}=2a$. For the chosen
$\omega_0$ this occurs approximately for $r_\textrm{i}= 0.09 a$.
Fig.~\ref{electric_field}(b) renders a similar plot to
Fig.~\ref{electric_field}(a) but, this time, all cylinders in the
interface monolayer have that particular $r_\textrm{i}$. The
presence of a surface mode being fed by the waveguide is clear in
Fig.~\ref{electric_field}(b), but as this mode does not couple to
radiative modes, the radiated power again spreads in all
directions as corresponds to an aperture size of the order of
$\lambda_0$. This point is corroborated by
Fig.~\ref{radiation_pattern} (red curve) showing an almost uniform
$S_r(\theta)$ for the case considered. In order to alter the
radiation pattern, the surface mode should couple to the continuum
of radiative modes. This can be achieved by introducing a periodic
modulation to the PC surface. In this way the surface state can
``pick up" a reciprocal lattice vector of the superimposed lattice
and connect to waves inside the light cone. Out of the different
possibilities for this modulation, we present here results for a
simple choice with a corrugation period $\Lambda=2a$: at either
side of the waveguide exit, the cylinders at the PC-vacuum
interface in the first $N$ even positions (we order the cylinders
consecutively according to their distance to the waveguide) are
shifted a distance $\Delta z$ along the $z$-axis.
Fig.~\ref{electric_field}(c) renders the real space electric field
amplitude for $\Delta z=-0.3 a$ and $N=9$. This plot is radically
different from the one obtained for systems without surface mode
or with a non-radiative surface mode, distinctly showing the
re-emission process. The beaming characteristics are clearly seen
in the corresponding far-field transmission angular distribution
shown in Fig.~\ref{radiation_pattern} for $N=1,3,6,9$. These
curves present a strong redistribution of the transmitted
radiation into the forward direction, with a beam angular width of
a few degrees.

In the corrugated metal case, a focus in the Fresnel regime was
found associated to the beaming in the far-field
region~\cite{FJAPL}. This is also the situation in the present
case. Fig.~\ref{axial_cuts}(a) shows the real space dependence of
the Poynting vector modulus, $S$, for a modulated surface with
$N=3$, where already the presence of a collimated beam and a focus
is apparent. In similar way to the metallic counterpart, the focus
characteristics depend on the number of periods of modulation in
the surface. As illustration, Fig.~\ref{axial_cuts}(b) shows
$S_z(z, x=0)$ for $N=0,1,3,6,9$. The maxima of these curves
provide the focal distance of the, very elongated, focus. In order
to further assign these effects (beaming and lensing) to the
existence of a leaky surface mode, and to rule out that they are
only due to the modification of the waveguide opening induced by
the corrugation, the $x$-component of the Poynting vector, $S_x$,
can be scrutinized along the interface. Fig.~\ref{interface_cuts}
shows $S_x(x, z=7.5a)$ for the cases of no surface mode, surface
mode ($N=0$), and surface leaky mode for $N=3,9$. For $N=0$, after
an initial dependence on $x$ for small distances to the waveguide
(due to the coupling between the waveguide mode and the surface
mode), the modal carried power remains constant, as expected for
the non-radiative surface mode present in the system. On the
contrary, cases with $N\neq 0$ show that the field decreases as
the mode travels along the surface, due to the radiation losses
responsible for the beaming and lensing phenomena. Notice that the
case with a surface termination that does not support surface
modes or resonances, presents an almost negligible intensity along
the surface. It is worth stressing here that, although we have
chosen a favorable range of parameters in order to show that the
cited effects exist, no attempt has been made in order to optimize
the structures. This is already apparent in
Figs.~\ref{electric_field}(b,~c), where the strong electric field
modulation inside the waveguide is due to a high reflection
[$R=52.3\%$ and $R=24.5\%$ in (b) and (c), respectively] caused by
the large impedance mismatch between waveguide and free space
regions. This reflection can be largely modified by very local
changes of the dielectric structure at the opening (such as
tapering~\cite{Moreno02b}). Similarly, the coupling between
waveguide mode and surface resonance surely has a strong influence
on the transmission angular distributions. Which geometry is
optimal is an interesting question that, however, falls outside
the scope of this paper.

Beaming at other angles is also possible. For instance, for the
chosen frequency $\omega_0$, shifting normally to the surface one
every three cylinders ($\Lambda=3a$), results in two collimated
beams at $\theta=\pm 24^{\circ}$. For the system presented in this
paper, coupling to a surface state with an in-plane wave vector
different from $2 \pi / \Lambda$ (which would occur at certain
frequencies different from $\omega_0$) would also result in
beaming at $\theta \ne 0^{\circ}$.

\begin{figure}[t]
\includegraphics{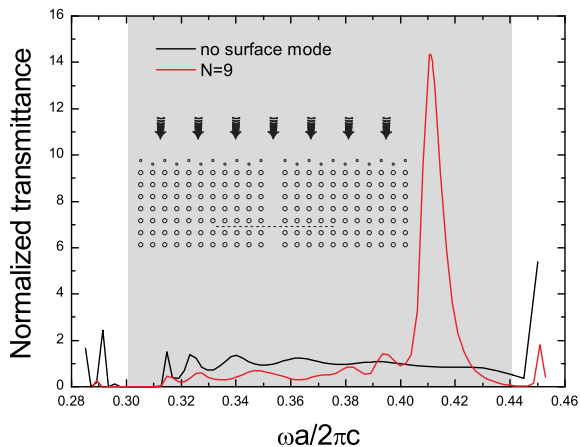}
\caption{\label{enhanced_transmission} Normalized transmittance
through a waveguide in a PC slab. The gray shading represents the
PBG. Black curve: standard interface (no surface mode supported);
red curve: interface with monolayer of cylinders with radius
$r_\textrm{i}= 0.09 a$ (one leaky surface mode supported) and
corrugated with $N=9$ indentations. The considered structure is
finite [slab thickness: 8 layers, lateral extension: $40 a$ (with
the central cylinder line removed to create the waveguide)]. The
employed normalization is explained in the text.}
\end{figure}

As previously said, another interesting effect found in metallic
structures is that a structure presenting beaming when corrugated
in the exit surface, is capable of presenting enhanced
transmission when that corrugation is induced in the input
surface. Fig.~\ref{enhanced_transmission} shows the transmittance
through a waveguide in a PC slab when illuminated by a normally
incident plane wave. The black curve corresponds to a regular
unstructured interface [same as in Fig.~\ref{electric_field}(a)],
whereas the red curve corresponds to the corrugated interface
shown in Fig.~\ref{electric_field}(c). The transmittance has been
normalized to the power that impinges onto a length $a$ (i.e.,
roughly the waveguide cross section). In both curves rendered in
Fig.~\ref{enhanced_transmission}, transmission outside the PBG is
due to the EM field permeating through the crystal, while the
region of transmission of order unity inside the gap is due to
propagation through the waveguide. Interestingly,
Fig.~\ref{enhanced_transmission} shows a large resonant
transmission inside the gap for the PC with a modulated surface.
The resonant frequency is the same that was previously obtained
for beaming of light out of a waveguide. Notice that, even in this
case without full optimization of the geometry, the energy flux
collected into the waveguide corresponds to the one impinging in a
cross section around $14a$. This effective cross section could be
increased, for instance, by considering a shallower corrugation
with a larger lateral size. Unlike the analog situation in
metallic structures, here the system is lossless, and therefore
dissipation does not set a limit to the possible device lateral
size.

To summarize, in this paper we have demonstrated the use of
surface leaky modes, appearing in PCs under appropriate
conditions, to collimate EM radiation out of PC waveguides. Very
narrow beams with angular distributions of a few degrees can be
achieved by modulating the PC exit surface. Furthermore, these
surface resonances may also be used for injecting light
efficiently into these waveguides. This enhanced transmission
property could have interesting applications in the efficient
coupling between conventional optical fibers and PC-based
waveguides. (Funding by the Swiss National Science Foundation/NCCR
Quantum Photonics and financial support by the Spanish MCyT under
contracts MAT2002-01534 and MAT2002-00139 are gratefully
acknowledged).


\end{document}